\documentclass[aps,prb,superscriptaddress,twocolumn,showpacs,amsmath,amssymb]{revtex4}
\usepackage{graphicx}
\usepackage{bm}

\newcommand{\Hamil}{{\cal H}}
\newcommand{\ket}[1]{|{#1}\rangle}
\newcommand{\bra}[1]{\langle{#1}|}

\newcommand{\av}[1]{\langle{#1}\rangle}

\newcommand{\com}[2]{[{#1},{#2}]}
\newcommand{\anticom}[2]{\{{#1},{#2}\}}

\newcommand{\ddtinline}{\partial/\partial t}

\def\Hameff{{\cal H}^\text{eff}}
\def\barhameff{\bar{\cal H}^\text{eff}}

\def\ket#1{|#1\rangle}
\def\bra#1{\langle#1|}

\def\bfI{{\bf I}}
\def\bfJ{{\bf J}}

\def\bfS{{\bf S}}

\def\tr#1{{\rm tr}\;#1}

\def\leade#1{\varepsilon_{#1\sigma}}
\def\dote#1{\varepsilon_{#1}}
\def\c#1{c_{#1\sigma}}
\def\cdagger#1{c_{#1\sigma}^{\dagger}}
\def\cs#1{c_{#1}}
\def\csdagger#1{c_{#1}^{\dagger}}

\def\up{\uparrow}
\def\down{\downarrow}

\def\im{{\rm Im}}

\newcommand{\beqa}{\begin{eqnarray}}
\newcommand{\eeqa}{\end{eqnarray}}

\begin{document}
 \title{ Nuclear Spin Noise and  STM Noise Spectroscopy}
 \author{A. V. Balatsky}
\email{avb@lanl.gov}  \affiliation{Theoretical Division, Los
Alamos National Laboratory, Los Alamos, New Mexico 87545, USA}
\author{J. Fransson} \email{Jonas.Fransson@fysik.uu.se}
\affiliation{Department of Materials Science and Engineering,
Royal Institute of Technology (KTH), SE-100 44\ \ Stockholm,
Sweden}
 \affiliation{Physics Department, Uppsala University, Box
530, SE-751 21\ \ Uppsala, Sweden}
\author{D. Mozyrsky}
\email{mozyrsky@lanl.gov} \affiliation{Theoretical Division, Los
Alamos National Laboratory, Los Alamos, New Mexico 87545, USA}
\author{Yishay Manassen}
\email{manassen@bgumail.bgu.ac.il}
 \affiliation{Department of Physics and the Ilse Katz Center
  for Nanometer Scale Science and Technology, Ben Gurion University, Beer Sheva, 84105, Israel}

\begin{abstract}
We consider fluctuations of the electronic spin due to coupling to nuclear spin. Noise spectroscopy of an electronic spin can be revealed in the Scanning Tunnelling Microscope (STM). We argue that the noise spectroscopy of electronic spin can reveal the nuclear spin dynamics due to hyperfine coupling. Tunnelling current develops satellites of the main lines at Larmor frequency and at zero frequency due to hyperfine coupling. We also address the role of the rf field that is at or near the resonance with the nuclear hyperfine field. This approach is similar to Electron Nuclear Double Resonance (ENDOR), in that is allows one to detect nuclear spin dynamics indirectly through its effect on electronic spin.
\end{abstract}
\maketitle

\section{Introduction}
\label{sec-introduction}
Noise spectroscopy is a technique that allows one to
measure spectroscopic properties by observing nontrivial
 features in the noise. The main feature of noise
 spectroscopy is that the noise spectrum has encoded
  in it spectroscopic features that correspond to
   physical excitations in the system, such as
   atomic levels, Zeeman split levels of electrons
    or nuclear levels. The transition between
    levels would cause enhanced dissipation
    when the energy transferred equals the
     energy difference between those levels. This enhanced
      dissipation also would imply enhanced fluctuations
       at the same frequencies/energies, as follows from the fluctuation-dissipation theorem. Experiments that prove utility of the noise are available from many fields. An incomplete list includes nuclear spin noise,\cite{Sleater85} Faraday rotation noise in the alkali atoms,\cite{Aleksandrov81,Mitsui01,Crooker04} and acoustic noise.\cite{Lobkis01} Recently noise spectroscopy was used to detect a single electronic spin with Magnetic Resonance Force Microscopy by an IBM group.\cite{Rugar04}

Therefore, in principle, noise measurement could be  a powerful tool to investigate the dynamics of the system. Sometimes noise is easier to measure and then noise spectroscopy could be even a preferred technique to investigate nano-scale systems.\cite{Rugar04,Crooker04}

One example of the noise spectroscopy relevant for us
here is Electronic Spin Resonance Scanning Tunneling
Microscopy, ESR-STM.  ESR-STM is a technique that is
 using the extremely local  nature of the STM measurement
 to detect the noisy precession of spin centers on the
 nonmagnetic surface. When a tip of an STM is located
  above a paramagnetic spin center the tunneling current
   is modulated by the precession in the presence of
   external field. It was shown\cite{22,23} that the
   ac current at the Larmor frequency is spatially
   localized within $0.5-1nm$. It is the spatial
   localization that suggests  that this technique
   is capable of detecting a single spin. In
   addition it was proved that  the frequency
    of the signal is dependent on real time on
     the size of the magnetic field.\cite{24,25}
      More recently  similar experiments have been
      done on the paramagnetic BDPA molecule.\cite{26}
       The interest in this technique has risen sharply
       recently,  due to the possibility to manipulate
        and detect a single spin\cite{27,33} and due
        to the possibility to use it for quantum
        computation.\cite{27,28}  There have been
        many proposals for the mechanism of this
        phenomenon.\cite{29,30,32,33,34,35,36}
        Present experiments are not sufficient
        to constraint possible mechanism and
        further investigation would help to elucidate the nature of the effect.

Recently Durkan reported the measurements
of the noise in  the ESR-STM on a TEMPO
 molecule.\cite{Durkan04} This molecule
 is well characterized, and it contains
 nitrogen $N$ with the nuclear spin $I=1$.
   ESR spectrum of TEMPO in the bulk is known
   to exhibit the hyperfine splitting on the
   order of ~ 15 G that corresponds to the
    free ($g_e = 2$) electron precession frequency
    on the order of 45 Mhz. The main new observation
     that is important in our context is that the ESR-STM on
      TEMPO molecule produced three peaks that possibly
      corresponds to the hyperfine split Larmor line in
       current spectrum. If reproduced, this observation
        opens up a new possibilities in noise spectroscopy in STM.

The purpose of this paper is to consider the case of
coupling of the electronic impurity spin $S$ to nuclear
spin $I$ via hyperfine coupling. We investigate the noise
spectroscopy of the nuclear spin and coupled electron-nuclear
 dynamics as it   might be seen in STM experiments. Indeed all
  the previous discussions of the mechanisms so far have focused
   on the dynamics of the electronic impurity spin. Hyperfine
    coupling to the nuclear spin would allow one to measure dynamics, relaxation times and transitions between the nuclear levels.

We find that the hyperfine coupling produces  additional
satellite lines in the noise spectra for the localized
 impurity spin that will be split away  from  Larmor
 line by the amount proportional to hyperfine coupling
  $A$. We also consider the case of the rf field at
  frequency $\omega \sim A$ applied to the system.
  This is the case of Electronic-Nuclead Double Resonance
   (ENDOR), where electron spin dynamics will be affected by
     the nuclear spin flips caused by the rf field. In the
      case of applied rf field we clearly pump energy
      into the system and hence measurement is not
      strictly a noise spectroscopy measurement.
      Still this is a set up most likely to be
      attempted experimentally and this is why
      we address it here as well.

The plan of the paper is as follows. We address the localized impurity spin susceptibility with the hyperfine coupling and the tunneling current modulations in the Sec.II.  We address a few specific cases, such as case of no rf field and case of no applied fields, neither dc nor ac. In all of these cases electronic spin will have nontrivial spectroscopic features, most notable the hyperfine split lines around Larmor line and around zero frequency. In Sec III we conclude with discussion of possible experiments.

\section{Probing the spin susceptibility via noise measurement}
\label{sec-body}
\begin{figure}[b]
\begin{center}
\includegraphics[width=8.5cm]{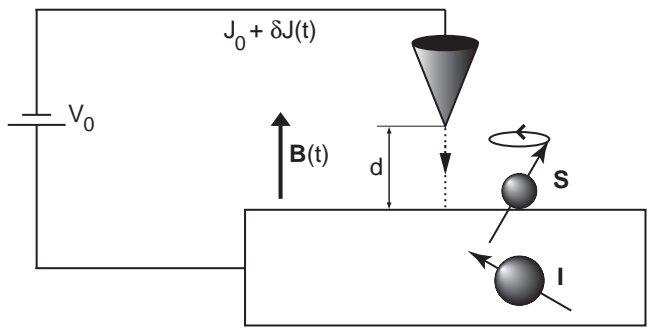}
\end{center}
\caption{Sketch of the experimental setup. The tip | surface separation distance is $d$. The single magnetic impurity atom with spin $\bfS$ on the substrate surface interacts with the nuclear spin $\bfI$ in the substrate.}
\label{fig-system}
\end{figure}
To be specific, consider the tunneling current between two contacts in the presence of a localised spin $\bfS$ interacting with a nuclear spin ${\bf I},\ I = 1$, to make a contact with the experiments on  TEMPO. The Hamiltonian of this system is written in the form
\begin{eqnarray}
\Hamil=\Hamil_L+\Hamil_R+\Hamil_S
    +\sum_{kp\alpha\beta}\csdagger{k\alpha}[t_0
        +t_1\bfS\cdot\sigma_{\alpha\beta}]\cs{p\beta},
\label{eq-Hamiltonian}
\end{eqnarray}
where $\Hamil_{L(R)}=\sum_{k(p)\sigma}\leade{k(p)}\cdagger{k(p)}\c{k(p)}$ models free electrons in the left (right), $L (R)$, lead, whereas
\begin{eqnarray}
\Hamil_S&=&B_0S^z+A\bfS\cdot\bfI+B_1I^x\cos{\omega t}
    +B_2S^x\cos{\omega t},
\label{eq-HS}
\\&&
    B_0\gg A\gg B_1,B_2
\nonumber
\end{eqnarray}
accounts for the interactions between the localised electronic and
nuclear spins. In the Hamiltonian, Eq. (\ref{eq-Hamiltonian}),
$\sigma_{\alpha\beta}$ denote the Pauli spin matrix vector with
matrix indices $\alpha,\beta$, the Fermionic
$\cdagger{k(p)},\c{k(p)}$ are creation and annihilation operators
of electrons in the $k$th ($p$th) eigenstate in the STM tip
(substrate surface), with spin $\sigma=\up,\down$. For simplicity
we have incorporated the electronic $g_e$ and nuclear gyromagnetic
ratio $g_N$ into the effective fields $B_0=g_eH_0,\ B_1=g_NH_1,
B_2=g_eH_1$ for the external dc field $H_0$ and ac field
$H_1\cos\omega t$.
\begin{figure}[t]
\vskip -1.5cm \rightskip 2cm 
\includegraphics[width=8.5cm, angle=90]{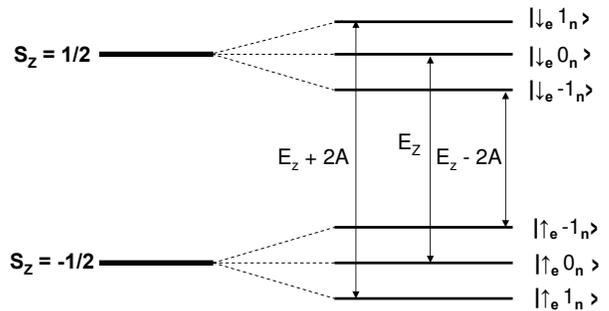}
\vskip -2.5cm \caption{The possible spin-flip transitions that may
occur in the system. The electronic spin-flips give rise to the
three major peaks in the transport current fluctuations spectrum
at the frequencies $E_z$ and $E_z\pm 2A$; see also Fig. 3 and Fig.
4.} \label{fig-trans}
\end{figure}
The last term in Eq. (\ref{eq-Hamiltonian}) describes tunneling of
the electrons from the tip into the substrate surface in the
presence of the localised electronic spin $\bfS$. This term only
give a contribution to the net steady state current by providing a
chemical potential shift (bias voltage drop) between the two bands
$\{\leade{k}\}$ and $\{\leade{p}\}$. In real systems, the hopping
matrix elements $\hat{t}=t_0+t_1\bfS\cdot\sigma$ have a $kp-$
dependence, which are omitted here in order to make the notation
more compact. The wavefunctions of our system are superpositions
of the direct product states
$\ket{\psi_L}\otimes\ket{\psi_S}\otimes\ket{\psi_R}$, e.g. the
direct product of the state of the STM tip, the impurity spin, and
the substrate surface. The tunneling matrix $\hat{t}$ in the last
term of Eq. (\ref{eq-Hamiltonian}) couples all of these different
states, of which the term proportional to $t_0$ describes the spin
independent tunneling while the term proportional to $t_1$
provides the spin dependent contributions arising from the
exchange interaction of tunneling electrons  to the magnetic atom.

Spin dependence of the tunneling arise due to direct
exchange dependence of the tunnel barrier.\cite{balatsky2002}
The overlap of the electronic wavefunctions of the tip and the
 surface, separated by a disctance $d$ is exponentially small
 and is given by a spin dependent tunneling matrix element
  $\hat{t}=\gamma\exp\{-\sqrt{(\Phi-J\bfS(t)\cdot\sigma)/\Phi_0}\}$,
  where the direct exchange between the tunneling electron
   spin $\sigma$ and the impurity spin $\bfS$ is explicitly
   included. Here, $J$ is the exchange interaction parameter
   between the electrons tunneling from the tip to the surface
    and the precessing impurity spin $\bfS$. The tunneling
    barrier height $\Phi$ is typically a few eV, whereas
    $\Phi_0=\hbar^2/(8md^2)$ is related to the distance
     between the tip and the surface.\cite{tersoff1993}
Since the exchange term in the exponent is small
compared to the barrier height, we may expand the
exponent in $JS$, which gives $\hat{t}=t_0+t_1\bfS(t)\cdot\sigma$.
Here, $t_0=\gamma\exp{(-\sqrt{\Phi/\Phi_0})}\cosh{[JS/(2\sqrt{\Phi\Phi_0})]}$
describes the spin independent tunneling,
while $t_1=\gamma\exp{(-\sqrt{\Phi/\Phi_0})}\sinh{[JS/(2\sqrt{\Phi\Phi_0})]}$ accounts for the spin dependent tunneling amplitude. For estimates we may employ the typical thumb rule $t_1/t_0\approx JS/(2\Phi)\ll1$.

We will assume that the tunneling electrons are partially spin polarised.
 This can be achieved in several situations. For example, in ferromagnetically
 coated tips a potential difference $2\delta\mu_\sigma$ separates the spin bands
  $\leade{k}=\dote{k}+\sigma\delta\mu_\sigma$. Ferromagntically ordered tips have
  proven successful in the study of magnetic structures.\cite{fmtips} Another
   approach to obtain a spin polarised current is to use an antiferromagntically
   coated tip with no ferromagnetic order.\cite{afmtips} Such tips have the benefit
    of a vanishing dipolar field and should, therefore, have a negligible influence
     on the precession frequency of the impurity spin. A ferromagnetically ordered
     tip may produce a field of ${\cal O}(1)$ T at a separation of few \AA ngstr\"oms from the
     surface. Such fields therefore leads to huge precession frequencies which are
     difficult to measure. In what follows, however, we will not be concerned of
     how the spin polarised current is generated. We define a parameter ${\cal A}$
     which relates the spin polarised current to the net tunneling current. This
     is the parameter that will be determined by a particular microscopic model
     of the tip. Thus, we will henceforth treat ${\cal A}$ as a phenomenological parameter.

For conciseness, we employ the Heisenberg picture for all operators $\{O\}$,
i.e. $O(t)=\exp{[iHt]}O_S\exp{[-iHt]}$, with $O_S=O(t)$ for operators in the
Schr\"odinger picture. All finite temperature expectation values $\av{O(t)}$
will represent $\sum_ip_i\bra{\psi_i(0)}O(t)\ket{\psi_i(0)}$, where $\psi_i(t=0)$
is the zero time wave function from the Schr\"odinger picture, whereas $p_i$ is
its probability within the density matrix formulation.

Now, for a qualitative description of the effect addressed here consider the
charge current $J(t)=-e\ddtinline N_L(t)$. Since we are considering the steady
state regime it is, by charge and current conservation, sufficient to consider
 the current in the tip (or in the substrate) only. By a direct calculation,
 using the Heisenberg equation of motion, we find that
\begin{equation}
J(t)=-2e\im\sum_{kp\alpha\beta}\Bigl\langle\csdagger{k\alpha}[t_0
        +t_1\bfS(t)\cdot\sigma_{\alpha\beta}]\cs{p\beta}\Bigr\rangle,
\label{eq-J1}
\end{equation}
where $e$ is the electronic charge. Hence, we see that the tunneling current
can be partitioned into a spin independent part $J_0(t)=-2e\im\sum_{kp\sigma}t_0\av{\cdagger{k}\c{p}}$ and a spin dependent part
\begin{equation}
\delta J(t)=\av{\delta\bfJ(t)}=et_1\av{\bfS(t)\cdot \bfJ_s(t)},
\label{eq-spinJ}
\end{equation}
which depends on the localised moment $\bfS(t)$, where
\begin{equation}
\bfJ_s(t)=-i\sum_{kp\alpha\beta}\csdagger{k\alpha}\sigma_{\alpha\beta}\cs{p\beta}
    +H.c.,
\label{eq-spinJop}
\end{equation}
is the spin dependent contribution to the tunneling current. The $z$-component
 of this expression, e.g. $J^z_s(t)=-i\sum_{kp}(\csdagger{k\up}\cs{p\up}-\csdagger{k\down}\cs{p\down})+H.c.$,
  describes the net flow of spin $\up$ and spin $\down$ carriers, whereas the transversal
  component $(J^x_s,J^y_s)(t)=-i\sum_{kp}(\csdagger{k\up}\cs{p\down}+\csdagger{k\down}\cs{p\up},-i\csdagger{k\up}\cs{p\down}+i\csdagger{k\down}\cs{p\up})+H.c.$
  accounts for spin flip transitions of the tunneling electrons caused by the interactions with the precessing impurity spin.

To the lowest order in the tunneling amplitude $t_1$, the electronic current-current
 correlation function arising due to the spin dependent part of the current is given by
\begin{eqnarray}
\av{\anticom{\delta\bfJ(t)}{\delta\bfJ(t')}}=
    (et_1)^2\av{S^i(t)S^j(t')}\av{J_s^i(t)J_s^j(t')}
\nonumber\\
    +(t\leftrightarrow t'),
\label{eq-IIcorrelation}
\end{eqnarray}
where $\av{\anticom{\cdot}{\cdot}}$ denotes the symmetrized correlator
whereas $i,j=x,y,z$ signify the spin components. Thus, to lowest non-trivial
order in $t_1$ we can treat the two temporal correlation functions
 $K^{ij}(t-t')=\av{S^i(t)S^j(t')}$ and $C(t-t')=\av{J_s^i(t)J_s^j(t')}\rightarrow\av{J_s^i(t)}\av{J_s^j(t')},\ (|t-t'|\rightarrow\infty)$
 independently. To make a connection with the main proposal of this paper we note
 that the Fourier transform of the symmetrized correlation function $\av{\anticom{\delta\bfJ(t)}{\delta\bfJ(t')}}$ is
 the current noise spectrum at various frequencies arising from the localised electronic spin. In Fourier space, the
  current noise spectrum is given by the convolution of the two power spectra associated with $\bfS$ and $\sigma$, e.g.
\begin{equation}
\av{|\delta\bfJ(\dote{})|^2}=\frac{(et_1)^2}{2\pi}\int
    K^{ij}(\dote{}')C(\dote{}-\dote{}')d\dote{}'+(\dote{}\rightarrow-\dote{}).
\label{eq-Jnoise}
\end{equation}

In order to see the effect of the interactions between the localised electronic ($\bfS$)
and nuclear ($\bfI$) spin in the rotating magnetic field we have to calculate $K^{+-}(\dote{})$
and $K^z(\dote{})$, for which the details are given in the appendix, see Eqs. (\ref{appeq-Kpm}) and (\ref{appeq-Kz}). The resulting expressions for the correlation functions are given by
\begin{widetext}
\begin{eqnarray}
K^{+-}(\tau)&\sim&\biggl(
    \frac{1}{4}\biggl[1+2\Bigl(\frac{B_1}{2\omega_1}\Bigr)^2
        \biggl(1+\Bigl(\frac{B_1}{2\omega_1}\Bigr)^2\biggr)
            +5\Bigl(\frac{\Delta\omega}{\omega_1}\Bigr)^4\biggr]
    +\frac{1}{4}\Bigl(\frac{B_1}{2\omega_1}\Bigr)^2
        \biggl(3-\Bigl(\frac{\Delta\omega}{\omega_1}\Bigr)^2\biggr)\cos\omega\tau
    +\frac{3}{4}\Bigl(\frac{B_1}{2\omega_1}\Bigr)\cos2\omega_1\tau
\nonumber\\&&
    +\frac{1}{8}\biggl(1+\frac{\Delta\omega}{\omega_1}\biggr)^4
        \cos(\omega+2\omega_1)\tau
    +\frac{1}{8}\biggl(1-\frac{\Delta\omega}{\omega_1}\biggr)^4
        \cos(\omega-2\omega_1)\tau
\nonumber\\&&
    +\Bigl(\frac{B_1}{2\omega_1}\Bigr)^2
        \biggl(1-3\Bigl(\frac{\Delta\omega}{\omega_1}\Bigr)^2\biggr)
            [\cos\frac{\omega\tau}{2}-\cos\omega_1\tau]
\nonumber\\&&
    +2\frac{\Delta\omega}{\omega_1}\Bigl(\frac{B_1}{2\omega_1}\Bigr)^2
        \biggl(1+\frac{\Delta\omega}{\omega_1}\biggr)
            \cos\frac{\omega+2\omega_1}{2}\tau
    -2\frac{\Delta\omega}{\omega_1}\Bigl(\frac{B_1}{2\omega_1}\Bigr)^2
        \biggl(1-\frac{\Delta\omega}{\omega_1}\biggr)
            \cos\frac{\omega-2\omega_1}{2}\tau
\nonumber\\&&
    +\frac{1}{2}\Bigl(\frac{B_1}{2\omega_1}\Bigr)^2
        \biggl(1+\frac{\Delta\omega}{\omega_1}\biggr)^2
            \cos(\omega+\omega_1)\tau
    +\frac{1}{2}\Bigl(\frac{B_1}{2\omega_1}\Bigr)^2
        \biggl(1-\frac{\Delta\omega}{\omega_1}\biggr)^2
            \cos(\omega-\omega_1)\tau
\nonumber\\&&
    -\frac{1}{2}\Bigl(\frac{B_1}{2\omega_1}\Bigr)^2
        \biggl(1+\frac{\Delta\omega}{\omega_1}\Bigr)^2
            \cos\frac{\omega+4\omega_1}{2}\tau
    -\frac{1}{2}\Bigl(\frac{B_1}{2\omega_1}\Bigr)^2
        \biggl(1-\frac{\Delta\omega}{\omega_1}\Bigr)^2
            \cos\frac{\omega-4\omega_1}{2}\tau
    \biggr)e^{-i2B_0\tau}.
\label{eq-Kpm}
\end{eqnarray}
and
\begin{eqnarray}
K^z(\tau)&\sim&
    \frac{1}{4}\Biggl[\biggl(1+\Bigl(\frac{\Delta\omega}{\omega_1}\Bigr)^2\biggr)
            \biggl(1+5\Bigl(\frac{\Delta\omega}{\omega_1}\Bigr)^2\biggr)
        +4\Bigl(\frac{B_1}{2\omega_1}\Bigr)^4\Biggr]
    +\frac{3}{4}\Bigl(\frac{B_1}{2\omega_1}\Bigr)^4
        [\cos\omega\tau+\cos2\omega_1\tau]
\nonumber\\&&
    +\frac{1}{8}\biggl(1+\frac{\Delta\omega}{\omega_1}\biggr)
        \biggl(\Bigl[1-\frac{\Delta\omega}{\omega_1}\Bigr]^2
            -2\Bigl(\frac{\Delta\omega}{\omega_1}\Bigr)^2\biggr)
                \cos(\omega+2\omega_1)\tau
\nonumber\\&&
    +\frac{1}{8}\biggl(1-\frac{\Delta\omega}{\omega_1}\biggr)
        \biggl(\Bigl[1+\frac{\Delta\omega}{\omega_1}\Bigr]^2
            -2\Bigl(\frac{\Delta\omega}{\omega_1}\Bigr)^2\biggr)
                \cos(\omega-2\omega_1)\tau
\nonumber\\&&
    +\Bigl(\frac{B_1}{2\omega_1}\Bigr)^2
        \biggl(1+3\Bigl(\frac{\Delta\omega}{\omega_1}\Bigr)^2\biggr)
            [\cos\frac{\omega\tau}{2}+\cos\omega_1\tau]
    -2\Bigl(\frac{\Delta\omega}{\omega_1}\Bigr)^2
        \Bigl(\frac{B_1}{2\omega_1}\Bigr)^2
            [\cos\frac{\omega+2\omega_1}{2}\tau
                +\cos\frac{\omega-2\omega_1}{2}\tau]
\nonumber\\&&
    -\frac{1}{2}\Bigl(\frac{B_1}{2\omega_1}\Bigr)^4
        [\cos(\omega+\omega_1)\tau+\cos(\omega-\omega_1)\tau
        +\cos\frac{\omega+4\omega_1}{2}\tau+\cos\frac{\omega-4\omega_1}{2}\tau]
\label{eq-Kz}
\end{eqnarray}
\end{widetext}
In the above equations we have defined the
detuning parameter $\Delta\omega=A-\omega/2$
and the parameter $\omega_1=\sqrt{\Delta\omega^2+(B_1/2)^2}$.
Resonant conditions of the system is
given for $\Delta\omega=0\ \Leftrightarrow\ \omega=2A$ giving $\omega_1=B_1/2$.
While the coefficients $(B_1/2\omega_1)^{2n}\ n=1,2$, rapidly decays to zero
out of resonance, the coefficients $(\Delta\omega/\omega_1)^{2n},\ n=1,2$,
rapidly grows to unity, since the amplitude of the rf field $B_1\ll A$.
Hence, all terms in Eqs. (\ref{eq-Kpm}) and (\ref{eq-Kz}), but the ones
proportional to $\cos(\omega\pm2\omega_1)\tau/2$, contribute to the
spectrum at resonance, whereas only
the peaks at $\dote{}=2B_0$ and $\dote{}=2B_0\pm(\omega\pm2\omega)$ give a
non-negligible contribution to the spectrum out of resonance. Below we
will focus on few experimentally relevant  possibilities.

\subsubsection{$B_0\neq0,\ B_1\neq0$}
\label{sec-B0AB1}
This case corresponds to applying an external DC field and rf field in or close to
nuclear resonance, which is known as Electronic Nuclear double Resonance
(ENDOR).\cite{Slichter} Under those circumstances the nuclear spin dynamics
is influenced by the rf field, which is reflected in the dynamics of the
electonic spin. This is perhaps one of the most relevant cases for nuclear
spin noise to be seen in STM experiments.

\begin{figure}[t]
\begin{center}
\includegraphics[width=8.5cm]{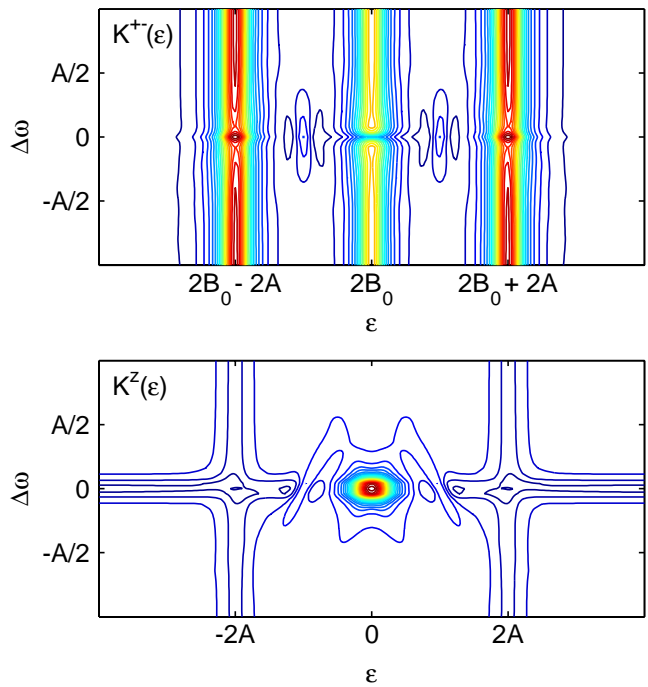}
\end{center}
\caption{Contour plots of the spectral intensities $K^{+-}(\epsilon)$ (upper panel)
and $K^z(\dote{})$ (lower panel) as function of the detuning $\Delta\omega$ in Fourier space ($\dote{}$). The spectra are generated by Lorentzian functions of uniform width $\Gamma/B_0=1/60$, using $A/B_0=1/12$ and $B_1/B_0=1/100$.}
\label{fig-K}
\end{figure}

Clearly, the correlation function $K^{+-}$ in Eq. (\ref{eq-Kpm}) has a central peak at the Larmor frequency  $\dote{}=2B_0$, which at resonance is represented by the first, third and seventh terms in Eq. (\ref{eq-Kpm}). This peak should also be measurable out of resonance since the constant (first) term is non-vanishing for all frequencies of the rf field.  There are side-bands around the frequency $\dote{}=2B_0\pm2A$, represented by the second, third, fourth, tenth, and eleventh terms, of which only the fourth and fifth are present out of resonance. Hence, also these peaks should be seen for all rf frequencies. However, at resonant conditions there are peaks around $\dote{}=2B_0\pm A$, represented by the sixth, eighth, ninth, twelfth, and thirteenth terms in Eq. (\ref{eq-Kpm}), which are not expected to be measurable appreciably far out of resonance. The same observations hold for the correlation function $K^z$, although its spectrum is centred around $\dote{}=0$. Hence, the central peak of this part is hidden in the white noise spectrum.

In Fig. \ref{fig-K} we display a contour plot of the correlation functions $K^{+-}$ (upper panel) and $K^z$ (lower panel) as function of the Fourier frequency $\dote{}$ and detuning $\Delta\omega$. Especially for $K^{+-}$ it is readily seen that there are five peaks ($2B_0,\ 2B_0\pm A,\ 2B_0\pm2A$) around resonant conditions $\Delta\omega=0$, while only three peaks remain out of resonance. In Fig. \ref{fig-Ksum} (solid) we show that total spectrum at resonance for the same conditions as in Fig. \ref{fig-K}.

\subsubsection{$B_0\neq0,\ B_1=0$}
\label{sec-B0A}
\begin{figure}[t]
\begin{center}
\includegraphics[width=8.5cm]{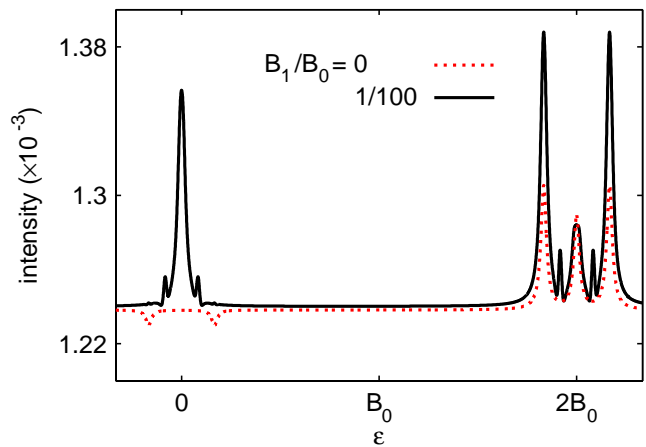}
\end{center}
\caption{Intensity of the normalised $K^{+-}(\dote{})+K^z(\dote{})$ with (solid) and without (dotted) rf field applied over the system.}
\label{fig-Ksum}
\end{figure}

The second case we consider is the case of a nuclear spin noise in a nondriven limit: $B_0\neq0,\ B_1=0\ (\omega=0)$.
 In this case the nuclear hyperfine field provides a flustuating field sampled by the electronic spin. Thus the total
 field will be given by a sum of the external field, $B_0$, and the $AI_z$ term. In this case the Larmor line
 will acquire sidebands due to the nuclear hyperfine field. This is clearly seen from Eq. (\ref{eq-Kpm}), which in the present case reduces to
\begin{equation}
K^{+-}(\tau)\sim\frac{5}{4}(1+8\cos2A\tau)e^{-i2B_0\tau},
\label{eq-Kpmnorf}
\end{equation}
since $\omega_1=|\Delta\omega|=A$ giving $\Delta\omega/\omega_1=1$. Obviously, this expression provides a main peak at $\dote{}=2B_0$ and sidebands at $\dote{}=2B_0\pm2A$, as is illustrated in Fig. \ref{fig-Ksum} (dotted). Similarly, the $z$-component of the spin-spin correlation function reduces to
\begin{equation}
K^z(\tau)\sim3-2\cos2A\tau.
\label{eq-Kznorf}
\end{equation}
Thus, there are dips at $\dote{}=\pm2A$, as seen in Fig. \ref{fig-Ksum}. Note that the amplitude of both $K^{+-}$ and $K^z$ are independent of the nuclear hyperfine field in this case, however the positions of the peaks obviously shift linearly with $A$.

Finally we comment on the case with no external field, e.g. $B_0,B_1=0$. Then, the impurity spin interact with the nuclear spin via the nuclear hyperfine field $A$. Assuming that the nuclear spin has a a very slow time dependent dynamics we can use the theory developed in the former section. Thus, under these circumstances the spin-spin correlation functions will reduce to
\begin{subequations}
\label{eq-KnoB0}
\begin{equation}
K^{+-}(\tau)\sim\frac{5}{4}(1+8\cos2A\tau),
\label{eq-KpmnoB0}
\end{equation}
\begin{equation}
K^z(\tau)\sim3-2\cos2A\tau.
\label{eq-KznoB0}
\end{equation}
\end{subequations}
As expected, the $z$-component of the correlation function is unaffected by
the absence of the field $H_0$, whereas the transverse component is
translated to $\dote{0}=0$ as $B_0\rightarrow0$. Hence, the
spectrum for $K^z$ is the same as in Fig. \ref{fig-Ksum} (dotted),
as well as the spectrum for $K^{+-}$, however, shifted to $\dote{}=0$.

\section{Conclusion}
\label{sec-conclusion}
In conclusion we presented a theory for nuclear spin
 noise spectroscopy. We considered a specific case of an ESR STM where
 the nuclear spin dynamics is revealed in the tunnelling current noise.
  We argue that noise spectroscopy is capable of detecting nuclear spin
  fluctuation  via a hyperfine coupling to localized
  impurity electronic spin. We find that the spectrum of the noise is
   rich and depends sensitively on the rf frequency and detuning,
   Fig. \ref{fig-K}. The main features of the spectrum are i) the
    peak at the Larmor electronic frequency that acquires hyperfine
    split satellites. This part of the spectrum is coming from the
     transverse spin fluctuations. The ENDOR like phenomenon occurs
      in the noise spectrum where the nuclear spin flips due to the
       rf field with nuclear resonance frequency directly affect
       the electronic spin dynamics. ii) the peak at zero frequency
        (that realistically will always be obscured by the $1/f$ noise)
         also acquires satellites at $\pm A$. This observation suggests
         that one measure the nuclear spin dynamics even in the absence
          of external fields. as long as the hyperfine lines are outside the $1/f$ noise peak.

\acknowledgements
This work was supported by DOE at Los Alamos. AVB is grateful to Karoly Holcer
for useful discussion and patient explanations. JF wants to acknowledge support from Carl Trygger's Foundation and Los Alamos National Laboratory.

\appendix
\section{Derivation of the spin susceptibility}
\label{app-derivation}
In the paper we have been considering the interaction between an impurity spin $\bfS\ (S=1/2)$ on the substrate surface and a nuclear spin $\bfI\ (I=1)$ in the substrate. The spin Hamiltonian for this system can be written as
\begin{equation}
\Hamil_S=B_0S_z+AS\cdot I+B_1I_x\cos{\omega t}+B_2S_z\cos{\omega t},
\label{eq-spinH}
\end{equation}
with $B_0\gg A\gg B_1,B_2$, which yields the effective Hamiltonian
\begin{equation}
\Hameff_S=B_0S_z+AS_zI_z+B_1I_x\cos{\omega t}.
\label{eq-effspinH}
\end{equation}

\subsection{$K^{+-}$}
\label{app-Kpm}
In order to see the effect of the electron spin flips on the susceptibility, we consider
the correlation function $K^{+-}(t_1,t_2)=\tr\{S_+(t_1)S(t_2)\rho\}$, where
 $S_\pm(t_1)=\exp\{i\int_0^{t_1}\Hameff_S dt\}S_\pm\exp\{-i\int_0^{t_1}\Hameff_S dt\}$, $S_+=\ket{\up}_e\bra{\down}$, $S_-=\ket{\down}_e\bra{\up}$, and $\rho=(\ket{\up}_e\bra{\up}+\ket{\down}_e\bra{\down})\otimes\openone_n$. Using the cyclic invariance under the trace gives
\begin{widetext}
\begin{eqnarray}
K^{+-}(t_1,t_2)&=&\tr_n\biggl\{\bra{\up}e^{-i\int_{t_1}^{t_2}\Hameff_S dt}\ket{\up}
    \bra{\down}e^{i\int_{t_1}^{t_2}\Hameff_S dt}\ket{\down}\biggr\}
    =e^{-i2B_0(t_2-t_1)}\tr_n\biggl\{\bra{\up}
        e^{-i\int_{t_1}^{t_2}\barhameff_S dt}\ket{\up}
        \bra{\down}e^{i\int_{t_1}^{t_2}\barhameff_S dt}\ket{\down}\biggr\}
\nonumber\\
    &=&e^{-i2B_0(t_2-t_1)}\tr_n\biggl\{
        Te^{-i\int_{t_1}^{t_2}(AI_z+B_1I_x\cos{\omega t})dt}
        Te^{i\int_{t_1}^{t_2}(-AI_z+B_1I_x\cos{\omega t})dt}\biggr\},
\end{eqnarray}
where $\barhameff_S=AS_zI_z+B_1I_x\cos{\omega t}$. Transforming the system into the rotating reference frame gives
\begin{eqnarray}
Te^{-i\int_{t_1}^{t_2}(AI_z\pm B_1I_x\cos{\omega t})dt}\rightarrow
    e^{-i\omega I_z\tau/2}
        e^{-i[\Delta\omega I_z/\omega_1\pm
            B_1I_x/(2\omega_1)]\omega_1\tau},
\end{eqnarray}
where the detuning parameter $\Delta\omega=A-\omega/2$, $\omega_1=\sqrt{\Delta\omega^2+(B_1/2)^2}$, and $\tau=t_2-t_1$ have been introduced. We introduce some notation by defining $Q_\pm=(\Delta\omega I_z\pm B_1I_x/2)/\omega_1$. In the algebra of the spin 1 operators we note that $I_z^{2n}=I_z^2$ for positive integers $n$ and $I_z^{2n+1}=I_z$ for all non-negative integers $n$. The same relations hold for $I_x$, i.e. $I_x^{2n}=I_x^2,\ n\geq1$, and $I_x^{2n+1}=I_x,\ n\geq0$. It is then easy to show that also $Q_\pm^{2n}=Q_\pm^2,\ n\geq1$, and $Q_\pm^{2n+1}=Q_\pm,\ n\geq0$. These rules for the algebra of the spin 1 operators give
\begin{equation}
e^{\pm i\alpha{\cal A}\tau}=1-{\cal A}^2(1-\cos\alpha\tau)\pm {\cal A}\sin\alpha\tau,
\end{equation}
where the operator ${\cal A}$ is either of $I_z,\ I_x$, or $Q_\pm$, and $\alpha$ is a scalar. Using these identities along with the facts that $\tr_nI_z=\tr_nI_x=\tr_nI_zI_x=\tr_nI_z^2I_x=\tr_nI_x^2I_z=0$, $\tr_nI_z^2I_x^2=1$, $\tr_nI_z^2=\tr_nI_x^2=2$, and $\tr_n\openone_n=3$, we find that the correlation function $K^{+-}(t_1,t_2)=K^{+-}(\tau)$ is given by (recall that the imaginary part of the trace vanishes)
\begin{eqnarray}
K^{+-}(\tau)&=&e^{-i2B_0\tau}\biggl\{
    \frac{1}{4}\biggl[1+2\Bigl(\frac{B_1}{2\omega_1}\Bigr)^2
        \biggl(1+\Bigl(\frac{B_1}{2\omega_1}\Bigr)^2\biggr)
            +5\Bigl(\frac{\Delta\omega}{\omega_1}\Bigr)^4\biggr]
    +\frac{1}{4}\Bigl(\frac{B_1}{2\omega_1}\Bigr)^2
        \biggl(3-\Bigl(\frac{\Delta\omega}{\omega_1}\Bigr)^2\biggr)\cos\omega\tau
\nonumber\\&&
    +\frac{3}{4}\Bigl(\frac{B_1}{2\omega_1}\Bigr)\cos2\omega_1\tau
    +\frac{1}{8}\biggl(1+\frac{\Delta\omega}{\omega_1}\biggr)^4
        \cos(\omega+2\omega_1)\tau
    +\frac{1}{8}\biggl(1-\frac{\Delta\omega}{\omega_1}\biggr)^4
        \cos(\omega-2\omega_1)\tau
\nonumber\\&&
    +\Bigl(\frac{B_1}{2\omega_1}\Bigr)^2
        \biggl(1-3\Bigl(\frac{\Delta\omega}{\omega_1}\Bigr)^2\biggr)
            [\cos\frac{\omega\tau}{2}-\cos\omega_1\tau]
\nonumber\\&&
    +\frac{1}{2}\Bigl(\frac{B_1}{2\omega_1}\Bigr)^2
        \biggl(1+\frac{\Delta\omega}{\omega_1}\biggr)^2
            \cos(\omega+\omega_1)\tau
    +\frac{1}{2}\Bigl(\frac{B_1}{2\omega_1}\Bigr)^2
        \biggl(1-\frac{\Delta\omega}{\omega_1}\biggr)^2
            \cos(\omega-\omega_1)\tau
\nonumber\\&&
    +2\frac{\Delta\omega}{\omega_1}\Bigl(\frac{B_1}{2\omega_1}\Bigr)^2
        \biggl(1+\frac{\Delta\omega}{\omega_1}\biggr)
            \cos\frac{\omega+2\omega_1}{2}\tau
    -2\frac{\Delta\omega}{\omega_1}\Bigl(\frac{B_1}{2\omega_1}\Bigr)^2
        \biggl(1-\frac{\Delta\omega}{\omega_1}\biggr)
            \cos\frac{\omega-2\omega_1}{2}\tau
\nonumber\\&&
    -\frac{1}{2}\Bigl(\frac{B_1}{2\omega_1}\Bigr)^2
        \biggl(1+\frac{\Delta\omega}{\omega_1}\Bigr)^2
            \cos\frac{\omega+4\omega_1}{2}\tau
    -\frac{1}{2}\Bigl(\frac{B_1}{2\omega_1}\Bigr)^2
        \biggl(1-\frac{\Delta\omega}{\omega_1}\Bigr)^2
            \cos\frac{\omega-4\omega_1}{2}\tau
    \biggr\}.
\label{appeq-Kpm}
\end{eqnarray}
By tuning into resonance, i.e. $\Delta\omega=0$ which is given at $\omega=2A$ and yields $\omega_1=B_1/2$, it follows that
\begin{eqnarray}
K^{+-}(\tau)&=&e^{-i2B_0\tau}\biggl\{
    \frac{5}{4}
    +\frac{3}{4}\cos2A\tau
    +\frac{3}{4}\cos B_1\tau
    +\frac{1}{8}\cos(2A+B_1)\tau
    +\frac{1}{8}\cos(2A-B_1)\tau
    +\cos A\tau-\cos\frac{B_1\tau}{2}
\nonumber\\&&
    +\frac{1}{2}\cos(2A+B_1/2)\tau
    +\frac{1}{2}\cos(2A-B_1/2)\tau
    -\frac{1}{2}\cos(A+B_1)\tau
    -\frac{1}{2}\cos(A-B_1)\tau
    \biggr\}.
\label{appeq-Kpmres}
\end{eqnarray}

\subsection{$K^z$}
\label{app-Kz}
By means of the same approach we derive the $z$-component of the spin susceptibility, $K^z(t_1,t_2)=\tr\{S_z(t_1)S_z(t_2)\rho\}$, where $S_z(t_1)=\exp\{i\int_0^{t_1}\Hameff_S dt\}S_z\exp\{-i\int_0^{t_1}\Hameff_S dt\}$ and $S_z=(\ket{\up}\bra{\up}-\ket{\down}\bra{\down})/2$. Using that $\com{S_z}{\Hameff_S}=0$ we find, for the $\ket{\up}\bra{\up}$-component of $\rho$,
\begin{eqnarray}
4K_z(t_1,t_2)&=&\tr_n\biggl\{\bra{\up}e^{-i\int_{t_1}^{t_2}\Hameff_Sdt}\ket{\up}
    \bra{\up}e^{i\int_{t_1}^{t_2}\Hameff_Sdt}\ket{\up}\biggr\}
    =\tr_n\biggl\{\bra{\up}e^{-i\int_{t_1}^{t_2}\barhameff_Sdt}\ket{\up}
    \bra{\up}e^{i\int_{t_1}^{t_2}\barhameff_Sdt}\ket{\up}\biggr\}
\nonumber\\&=&
    \tr_n\biggl\{Te^{-i\int_{t_1}^{t_2}(AI_z+B_1I_x\cos\omega t)dt}
        Te^{i\int_{t_1}^{t_2}(AI_z+B_1I_x\cos\omega t)dt}\biggr\}
    =\tr_n\biggl\{e^{-iI_z\omega\tau/2}e^{-iQ_+\omega_1\tau}
            e^{iI_z\omega\tau/2}e^{iQ_+\omega_1\tau}\biggr\}
\nonumber\\&=&
    \frac{1}{4}\biggl(1+\Bigl(\frac{\Delta\omega}{\omega_1}\Bigr)^2\biggr)
            \biggl(1+5\Bigl(\frac{\Delta\omega}{\omega_1}\Bigr)^2\biggr)
        +\Bigl(\frac{B_1}{2\omega_1}\Bigr)^4
    +\frac{3}{4}\Bigl(\frac{B_1}{2\omega_1}\Bigr)^4
        [\cos\omega\tau+\cos2\omega_1\tau]
\nonumber\\&&
    +\frac{1}{8}\biggl(1+\frac{\Delta\omega}{\omega_1}\biggr)
        \biggl(\Bigl[1-\frac{\Delta\omega}{\omega_1}\Bigr]^2
            -2\Bigl(\frac{\Delta\omega}{\omega_1}\Bigr)^2\biggr)
                \cos(\omega+2\omega_1)\tau
\nonumber\\&&
    +\frac{1}{8}\biggl(1-\frac{\Delta\omega}{\omega_1}\biggr)
        \biggl(\Bigl[1+\frac{\Delta\omega}{\omega_1}\Bigr]^2
            -2\Bigl(\frac{\Delta\omega}{\omega_1}\Bigr)^2\biggr)
                \cos(\omega-2\omega_1)\tau
\nonumber\\&&
    +\Bigl(\frac{B_1}{2\omega_1}\Bigr)^2
        \biggl(1+3\Bigl(\frac{\Delta\omega}{\omega_1}\Bigr)^2\biggr)
            [\cos\frac{\omega\tau}{2}+\cos\omega_1\tau]
    -2\Bigl(\frac{\Delta\omega}{\omega_1}\Bigr)^2
        \Bigl(\frac{B_1}{2\omega_1}\Bigr)^2
            [\cos\frac{\omega+2\omega_1}{2}\tau
                +\cos\frac{\omega-2\omega_1}{2}\tau]
\nonumber\\&&
    -\frac{1}{2}\Bigl(\frac{B_1}{2\omega_1}\Bigr)^4
        [\cos(\omega+\omega_1)\tau+\cos(\omega-\omega_1)\tau
        +\cos\frac{\omega+4\omega_1}{2}\tau+\cos\frac{\omega-4\omega_1}{2}\tau].
\label{appeq-Kz}
\end{eqnarray}
At resonance $\Delta\omega=0$ we then have
\begin{eqnarray}
K^z(\tau)_{\Delta\omega=0}&=&
    \frac{1}{8}\Bigl(
    10+6[\cos\omega\tau+\cos2\omega_1\tau]
    +\cos(\omega+2\omega_1)\tau+\cos(\omega-2\omega_1)\tau
    +8[\cos\frac{\omega\tau}{2}+\cos\omega_1\tau]
\nonumber\\&&
    -4[\cos(\omega+\omega_1)\tau+\cos(\omega-\omega_1)\tau
        +\cos\frac{\omega+4\omega_1}{2}\tau+\cos\frac{\omega-4\omega_1}{2}\tau]
        \Bigr)
\label{appeq-Kzres}
\end{eqnarray}

\end{widetext}

\end{document}